\documentstyle[aps,prl,preprint]{revtex}

\begin{document}

\draft

\title{Chaotic Scattering with Resonance Enhancement}
\author{B. Elattari\cite{brahim}, V. Kagalovsky, and H.A. Weidenm\"uller}
\address{Max-Planck-Institut f\"ur Kernphysik, 69029 Heidelberg, Germany}
\date{\today}
\maketitle

\begin{abstract}The passage of light or of electrons through a
  disordered medium is modified in the presence of resonances. We
  describe a simple model for this problem, and present first results. 
\end{abstract}

\section{Introduction and Motivation}During the last decade, many 
phenomena related to the passage of waves through a disordered medium
have been intensely studied. This holds both for electrons, i.e. for
the amplitude waves of the Schr\"odinger equation, and for light,
i.e. for classical electromagnetic waves. As examples, we mention
conductance fluctuations in mesoscopic probes, and weak and strong
localization effects including the enhanced backscattering of light
reflected from a disordered medium. For electrons, disorder is
typically caused by scattering from impurities. For light, disorder 
is often produced artificially: A powder is immersed into some
liquid. Random scattering occurs if the indices of refraction of 
powder and liquid are sufficiently different.  

The present paper has been motivated by an experiment performed several
years ago \cite{Alb91} on the scattering of light by a disordered
medium. In this experiment, the powder used (TiO$_2$) consisted of 
grains with a size distribution centered around a diameter of $220$
nm. For such granules, a Mie resonance occurs close to the wavelength 
$\lambda \sim 630$ nm of the laser light used in the experiment. This 
led to a resonance enhancement of the scattering. The effects of such 
enhancement were seen by comparing the diffusion constant $D$ 
(determined from the intensity autocorrelation function versus
frequency of the transmitted light), and the transport mean free
path $l$ (determined either from weak localization, i.e. from enhanced
backscattering, or from the dependence of the transmission intensity on
the length $L$ of the disordered slab). One expects $D$ and $l$ to
obey the relation $D = 1/3 \ v_E l$, with $v_E$ the energy transport 
velocity through the disordered medium. In the present case, the 
ratio $3 D / l$ yielded a value $v_E = (5 \pm 1) \ 10^7$  m s$^{-1}$ 
for the transport velocity $v_E$ which is about an order of magnitude
smaller than the phase velocity. This surprising result has been 
understood both for small concentration \cite{Alb91} and, more
recently, also for strong concentration \cite{Bus95} of the scatterers. 
Qualitatively speaking, the transport velocity is reduced because on 
its way through the medium, the energy is stored for some time in the 
Mie resonances. 

In this paper, we study the passage of waves through a medium with 
randomly positioned resonant scatterers from the point of view of 
random matrix theory (RMT). To the best of our knowledge, this has not
been done before. Our motivation is the following. First, the approach
taken in refs. \cite{Alb91} and \cite{Bus95} is not entirely
systematic. Indeed, it combines results of diagrammatic impurity 
perturbation theory (used for analysing the intensity autocorrelation 
function) with the use of the Bethe--Salpeter equation \cite{Alb91}, 
or of a mean--field approximation \cite{Bus95}, for the calculation of
the transport velocity $v_E$. It is our hope that a suitable 
modification of RMT (capable of accounting for the presence of 
resonance scattering) may yield a uniform account for both, average 
and fluctuation properties. This is all the more desirable since
during the last decade, RMT has developed into one of the most
important tools for investigating wave propagation in disordered
media. The problem studied in this paper calls for an extension of
existing RMT and would thus widen the scope of this theory. Second, 
RMT would make it conversely possible to put the present problem
into a wider context, and to study, for instance, the effect of
resonances on localization properties. 

We have embarked on this project only recently. The present paper is,
therefore, no more than a work--in--progress report. We simplify the
problem by focussing on scalar waves with a simple dispersion
relation (i.e., electrons), thus postponing the description of the 
propagation of light waves to a later stage. The RMT model with 
resonances is introduced 
in section \ref{M}. In section \ref{S}, Efetov's supersymmetric 
generating functionals \cite{Efe83} in the version of ref. \cite{Ver85} 
are used to handle the technical problems resulting from the model. 
First results obtained in the simplest case are presented in section 
\ref{R}. 

\section{The model}
\label{M}

To motivate the model, and to identify the relevant parameters, we
begin with a few qualitative considerations. We consider a volume $V$ 
of liquid containing a volume concentration $\mu$ (defined as volume of 
grains per total volume) of resonant powder. To simplify the problem, we 
assume that all grains (i.e., all Mie scatteres) are exactly equal, 
each having volume $\Phi$, and resonance frequency
$\omega_1$. Then, the number $m$ of grains in $V$ is $m = \mu V /
\Phi$. The total volume $m \Phi = \mu V$ occupied by the grains is 
denoted by $V_{grain}$, the remaining volume $V_{rem}$ left for the 
liquid is $V_{rem}=V(1 - \mu)$.

We denote by $ \rho_L^{(0)}( \omega)$ the density of states per unit
volume {\it in the liquid} for a classical scalar wave of frequency 
$\omega$. We have \cite{She95} 
\begin{equation}
\rho_L^{(0)}(\omega) = {\omega^2 \over 2 \pi^2 v_0^3 \hbar},  
\label{r}
\end{equation}
where $v_0$ is the phase velocity of the wave in the liquid. For 
$\mu = 0$, the total density of states in the liquid is 
$\rho_L = \rho_L^{(0)}(\omega) V$. For $\mu \neq 0$ we only 
have the remaining volume $V_{rem}$ at our disposal, so that 
$\rho_{L,rem}=\rho_L^{0} V_{rem} = \rho_L^{0} V (1 - \mu)$. For later 
use, we note that for a vessel of length $L = 17 \ \mu$m and area 
$A = 6 \ 10^3 \ (\mu$ m$)^2$, for $v_0 = c$ (the velocity of light 
in vacuum), and for a wave length $\lambda = 630$ nm, the mean level 
spacing $d_L = \rho_L^{-1}$ is about $4 \ 10^{-7}$ eV.  

With increasing concentration of scatterers, the remaining volume
shrinks, and with it the level density of the liquid. We may ask: 
Where is the ``missing level density'' $\rho_L^{0} V \mu $? To answer
this question, we consider a single grain. In this grain, there
are many Mie resonances, with a level density $\rho_{Mie}$ which is 
proportional to $\Phi$ (so that $\rho_{Mie} = \rho_{Mie}^{0} \Phi 
\ll \rho_{L})$. The total level density in all $m$ grains is
$m\rho_{Mie}=\rho_{Mie}^{0} m \Phi=\rho_{Mie}^{0} V \mu $. This is
exactly the missing contribution provided we have $\rho_{L}^{0} = 
\rho_{Mie}^{0}$. In practice, this condition will not hold, because 
the level density depends on the index of refraction of the medium. 
For purposes of the present qualitative discussion, we disregard 
this fact.

We see that the immersion of the grains in the liquid causes a 
{\it concentration of level density} at the position of the Mie 
resonances: While $\rho_{L}$ is lowered from $\rho_{L}^{0} V $ to 
$\rho_{L}^{0} V (1-\mu )$, $\rho_{Mie}$ increases from zero to 
$\rho_{Mie}^{0} V \mu$. We note that on the scale of the mean level
spacing $d_L = 1/(\rho_{L}^{0} V (1-\mu))$ in the liquid, 
$\rho_{Mie}$ is not smooth but sharply peaked at the 
position of the Mie resonances. The area under each peak is $m$. We
also note that typically, $m\gg 1$: For a concentration of $1$ 
percent, a grain size of $200$ nm, and a volume of $10^{-13}$
m$^3$, we have $m \sim 10^5$ \ !

We expect the shape of the peak of $\rho_{Mie}$ to be Lorentzian, with
a width $\Gamma$ given by the decay width of the individual Mie
resonances (we assume that the concentration is sufficiently low
so that the resonances do not ``talk'' to each other). Noticeable 
optical effects can be expected only if the peak height $ 2 m / 
(\pi\Gamma)$ is at least comparable with $\rho_L$, i.e. for $m d_L 
\sim \Gamma$, or for $\mu \geq \rho_L^{(0)} \Gamma$. When is this 
condition met? For light, we typically have $\Gamma \sim 0.1$ eV. 
Using this figure and the values for $\rho_L$ and $\Phi$ given above, 
we find $\mu \geq 2$ percent.

Naturally, these considerations are only semiquantitative. Nonetheless, 
they establish the following picture. For grain  concentrations in the
percent region, there occurs a significant local enhancement of the 
level density at the characteristic frequency $\omega_1$ of the Mie
scatterers, with a width given by the decay width $\Gamma$ of the Mie
resonance. Typically, this width is very much larger than the mean
level spacing $d_L$ in the liquid. For frequencies $\omega$ differing
from the Mie resonance frequency $\omega_1$ by more than a few units
of $\Gamma / \hbar$, the wave incident on a grain does not excite the
Mie resonance, and the scattering by the wave on the grains is a
totally random process kin to diffusive impurity scattering. As
$\omega$ approaches $\omega_1$, resonant scattering will be superposed
onto this random process. Even for $\omega = \omega_1$, there will
still be random scattering since upon hitting a Mie scatterer, only 
part of the incident wave will excite the resonance, the remainder
undergoing hard--sphere reflection. 

How can we model this situation in the framework of RMT? We recall how
diffusive scattering in the {\it absence} of Mie resonances is modelled
in the framework of RMT, for a quasi one--dimensional system of length
$L$ \cite{Iid90}: The system is considered as consisting of many
longitudinal slices of length $l \ll L$. Within each slice, the
Hamiltonian is modelled as a random Hamiltonian matrix (a member of
the random--matrix ensemble with the proper symmetry). Neighboring
slices are coupled by Gaussian--distributed uncorrelated random matrix
elements. The first and the last slice are coupled to the leads or,
equivalently, to the channels. Elements of the scattering matrix
connecting incident and outgoing channels define the conductance,
i.e. the transmission through the disordered region. 

With proper modifications, we use this model also for the present
problem. To begin with, we consider only Schr\"odinger waves. The
modifications caused by the different form of the Hamiltonian for
light waves will be addressed at a later stage. We retain the idea of 
a division of the quasi one--dimensional system into slices. To
account for the presence of Mie scatterers, the form of the
Hamiltonian has to be modified: Within each slice,
our Hamiltonian $\bf H$ is a matrix of dimension $N + m_S$ where 
$m_S$ is the number of Mie scatterers within each slice. The 
Hamiltonian $\bf H$ contains a purely random part (a matrix 
$H_{GOE}$ of dimension $N$ belonging to the Gaussian orthogonal 
ensemble [GOE]), a diagonal part of dimension $m_S$ with diagonal
elements $E_1 = \hbar \omega_1$ corresponding to the presence of $m_S$
Mie scatterers with equal resonance frequencies $\omega_1$, and a
rectangular matrix $V$ which couples the $m_S$ resonaces to
$H_{GOE}$. Since we will focus on the Hamiltonian for a single slice
in all that follows, we replace $m_S$ by $m$ for simplicity. With 
$I_m$ the $m$--dimensional unit matrix and with $T$ denoting the 
transpose, we thus have
\begin{equation}
{\bf H}=\left( \begin{array}{cc} \begin{array}[b]{c}
H_{GOE} \end{array} & V \\
V^T & E_1\times I_m
\end{array}
\right). 
\label{ham}
\end{equation}
The dimension $N$ of $H_{GOE}$ is taken to infinity at the end of the
calculation. The variance of the matrix elements $H_{GOE}$ is 
normalized as usual: The ensemble average of $(H_{GOE})_{\mu \nu} 
(H_{GOE})_{\mu' \nu'}$ is given by $(\lambda^2/N) 
(\delta_{\mu, \mu'} \delta_{\nu, \nu'} + 
\delta_{\mu, \nu'} \delta_{\nu, \mu'})$. Here, $\lambda$ (not to be
confused with the wave length of the scattered light) has the
dimension of an energy and determines the average level spacing
$d_S$ in each slice. For $m = 0$, the average level density obeys the
semicircle law, and the dependence of $d_S$ on energy $E$ is given 
by $\pi \lambda / N \sqrt{1 - (E/(2 \lambda))^2}$. For $m \neq 0$, we
expect the average level density to deviate from the semicircle law by
aquiring an additional peak at energy $E_1$. The width of the peak
will be determined by the matrix $V$. By assuming the 
$m$--dimensional matrix in the lower 
right--hand block to be diagonal, we implement the assumption that 
the $m$ resonances do not ``talk'' to each other, i.e., are not 
coupled directly. By the same token, we assume that resonances in 
different slices are not directly coupled to each other. Both 
assumptions are obviously valid only for a sufficiently low 
concentration of Mie scatterers. Because of this assumption, the 
rectangular matrix $V$ can, without loss of generality, be taken to 
be diagonal. Indeed, $V$ can be written in the form $O_1 V_D O_2^T$ 
where $V_D$ is diagonal, and where $O_1$ and
$O_2$ are orthogonal matrices of dimension $N$ and $m$, respectively. 
Transforming $\bf H$ with the orthogonal matrix $O = $ diag$(O_1,O_2)$,
and absorbing the matrix $O_1$ into $H_{GOE}$ by virtue of the
orthogonal invariance of the GOE, we obtain a new Hamiltonian of the
form of Eq.~(\ref{ham}), but with $V$ replaced by the diagonal matrix
$V_D$. Since we assume all scatterers to be identical, we take $V_D$ 
to be a  multiple of the unit matrix, $V_D = v \times I_m$. This
completes the definition of the model.

\section{Supersymmetric Formalism}
\label{S}

So far, we have only considered the simplest apects of the model
defined in section \ref{M}: For technical reasons, we have replaced
the GOE by the unitary ensemble (GUE), and we have calculated the mean
level density and the two--point level correlation function. This was
done to make sure that the model behaves as we expect, and to gain
experience with a Hamiltonian of the form of Eq.~(\ref{ham}). We have
used the supersymmetry technique of refs. \cite{Efe83,Ver85}. We omit
all details, use the notation of ref. \cite{Ver85}, and give only a 
few results which display characteristic differences between the usual
procedure, and the present model.

After averaging and the Hubbard--Stratonovich transformation, the
exponent of the generating functional for the two--level correlation
function (with energies $E \pm \omega$) has the form  
\begin{equation}
-\frac{N}{2\lambda}{\rm trg} \sigma^2 - {\rm trg} \log\left(\begin{array}{ccc}
E-\lambda\sigma - \omega \Lambda - J & 
0 & -V \\ 0 & E-\lambda\sigma  - \omega \Lambda - J & 0 \\ -V & 0 & E-E_1 -
\omega \Lambda - J \end{array}\right)
\begin{array}{l} \}m \\ \}N-m \\ \}m \end{array}.
\label{fun}
\end{equation}
The symbols $J$ stand for the source terms. The dimensions of the
block matrices are indicated at the side, and $V$ stands for $v \times
I_m$. Variation with respect to $\sigma$ leads to the saddle-point 
equation
\begin{equation}
\frac{\sigma}{\lambda}=(1-\frac{m}{N})\frac{1}{E-\lambda\sigma}+\frac{m}{N}
\frac{E-E_1}{(E-
\lambda\sigma )(E-E_1)-v^2}.
\label{sp}
\end{equation}
As usual, we have assumed that $\omega \sim 1/N$, and we have omitted
the (Infinitesimal) source terms. We take the limit $N \rightarrow
\infty$ {\it and keep $m$ fixed}. This yields the standard
saddle-point equation
\begin{equation}
\frac{\sigma}{\lambda}=\frac{1}{E-\lambda\sigma}
\label{ssp}
\end{equation}
with solution 
\begin{equation}
\sigma =\frac{E}{2\lambda}+i\Delta ,
\label{sol}
\end{equation}
where $\Delta =\sqrt{1-(E/2\lambda )^2}$. In contrast to the usual
case, the average level density is no longer given by the
saddle--point solution of Eq.~(\ref{sol}) but attains an extra piece
which is due to the last block in the matrix displayed in
Eq.~(\ref{fun}). Likewise, the two--point function differs from the
regular form, and has to be calculated by keeping carefully track of
the extra terms appearing in Eq.~(\ref{fun}). The results are as
follows. 

For the average density of states, we define the level density for the 
Hamiltonian ${\bf H}$ as $\rho(E) = \sum_i \delta(E - E_i)$. Let a bar
denote the ensemble average. Then, we find
\begin{equation}
\overline{\rho (E)} = \frac{\Delta (E)N}{\lambda\pi}+
m\frac{\Gamma\Delta (E)/\pi}
{(E-E_1-(E/2\lambda )\Gamma )^2+(\Gamma\Delta )^2},
\label{dos}
\end{equation}
where $\Gamma = v^2 / \lambda$ defines the effective width $\Gamma
\Delta$ of the Lorentzian. The form of Eq.~(\ref{dos}) is expected. 
Indeed, $\pi \lambda / (N \Delta)$ is the mean level spacing $d$
of the GUE and is to be identified with $d_L$, the mean level spacing
in the liquid. The last term in Eq.~(\ref{dos}) has area $m$, peak
height $m / (\pi \Gamma \Delta)$, and Lorentzian shape, all as 
anticipated in section \ref{M}. The occurrence of the factor 
$\Delta$ reflects the semicircle law for the GUE level density, 
and the factor $E / (2 \lambda)$ multiplying $\Gamma$ is an 
associated level shift which vanishes in the center of the semicircle 
where $E = 0$. The form of $\Gamma$ is also in keeping with
expectations. Indeed, we recall that the GUE matrix elements have
variance $\lambda^2 / N$. We expect the matrix elements $v$
occurring in Eq.~(\ref{sp}) to scale similarly with $N$, so that $v^2
= \alpha \lambda^2 / N$ where $\alpha$ is independent of $N$. Then,
$\Gamma = \alpha \lambda / N $ so that $\Gamma$ is given as $\alpha
\pi / \Delta$ times the mean level spacing of the GUE. For light
scattering, $\alpha \gg 1$ although we also consider the case $\alpha
\sim 1$ in this paper.

The Lorentzian in Eq.~(\ref{dos}) is plainly a superposition of $m$
individual resonances. It may come as a surprise that the indirect 
interaction between resonances caused by their joint coupling to the 
random Hamiltonian does not cause any level repulsion between the $m$
Lorentzians. To clarify this issue, we have solved the saddle--point 
equation (\ref{sp}) to next order in $N^{-1}$. We find that this leads
to an energy--dependent broadening of the Lorentzian in
Eq.~(\ref{dos}) which obviously is of order $N^{-1}$ relative to the 
terms we have kept. We conclude that within the approximations leading
to Eq.~(\ref{dos}), the indirect coupling between 
resonances is negligible. 

Since the presence of the $m$ resonances at energy $E_1$ destroys
translational symmetry, the evaluation of the level--level correlation
function is slightly more complicated than is the case usually. We
calculate
\begin{equation}
C(E,\omega )= \overline{\rho(E + \omega) \rho(E - \omega)} -
\overline{\rho(E + \omega)} \  \overline{\rho(E - \omega)}.
\label{cor}
\end{equation}
The disconnected terms cancel as usually if the ``boundary terms''
\cite{Hal95} or ``Efetov--Wegner'' terms \cite{Ver85} are taken into 
account. The remaining integrand contains an exponential with argument 
\begin{equation}
-2i\pi \omega/d \ {\rm trg} (1 + \alpha_1) + m \ {\rm trg} \ \log \left( (E -
E_1 - E \Gamma / (2 \lambda))^2 + \Delta^2 \Gamma^2 - \omega^2  - 2i
\omega \Gamma \Delta (1 + \alpha_1) \right). 
\label{int}
\end{equation}
Here, $\alpha_1$ is defined as in ref. \cite{Ver85}. It is easily 
seen that for $\omega \ll \Gamma$, this expression reduces to $-2i\pi
\omega / d_{eff}(E) \ {\rm trg} (1 + \alpha_1)$ where $d_{eff}(E)$ is the
inverse of the average level density $\overline{\rho(E)}$ given in
Eq.~(\ref{dos}). By virtue of the integration over the variables in
$\alpha_1$, the $\omega$--dependence of $C(E,\omega)$ is effectively cut 
off at $\omega \sim d_{eff}$. For values of $\Gamma \gg d_{eff}$, the 
normalized two--point autocorrelation function therefore essentially 
coincides with the one for the pure GUE {\it except for a rescaling 
of the average level spacing}. Differences can occur only in the wings
of the distribution. or in cases where $\Gamma \sim d_{eff}$. The 
figures presented in the next section are based on the exact
form of $C(E,\omega )$ given by
\begin{eqnarray}
&&C(E,\omega )= 
\overline{\rho(E + \omega) \rho(E - \omega)} =
{\rm Re}\left[
\frac{1}{2d^2}\left(\frac{iD}{2\pi\omega}\right)^{-m+1}\exp 
(2\pi\omega D/d)
\Gamma (-m+1,2\pi\omega (D-i)/d) 
\right.
\nonumber \\ 
&&\left.\times\int_{-1}^{1}d 
\lambda_2(iD+\lambda_2)^m 
\exp (-2i\pi\omega\lambda_2/d)
- -\frac{m}{2\pi d}\left(\frac{iD}{2\pi\omega}\right)^{-m}\exp (2\pi\omega D/d)
\left(\frac{D}{\omega}+\frac{1}{\Gamma\Delta}\right) 
\right.
\nonumber \\ 
&&\left.\times \Gamma (-m,2\pi\omega (D-i)/d)
\int_{-1}^{1}d
\lambda_2(iD+\lambda_2)^{m-1}\exp (-2i\pi\omega\lambda_2/d)
- -\frac{m^2}{8\pi^2}\left(\frac{iD}{2\pi\omega}\right)^{-m-1}
\frac{D^2+1}{\omega^2} 
\right.
\nonumber \\ 
&&\left.\times\exp (2\pi\omega D/d)
\Gamma (-m-1,2\pi\omega (D-i)/d)
\int_{-1}^{1}d
\lambda_2(iD+\lambda_2)^{m-2}\exp (-2i\pi\omega\lambda_2/d) \right],
\label{dosdos}
\end{eqnarray}
where $\Gamma(a,b)$ is the Gamma function, and where 
\begin{equation}
D = [(E-E_1-(E/2\lambda )\Gamma )^2+(\Gamma\Delta)^2-\omega^2]/
(2\omega\Gamma\Delta ). 
\label{d}
\end{equation}
The remaining integrals in Eq.~(\ref{dosdos}) can be calculated
analytically for every $m$. The resulting expressions are too lengthy
to be given here.

In the sequel, we also consider the normalized quantity
\begin{equation}
C_{nor}(E,\omega )=\frac{C(E,\omega)}{\rho (E-\omega )\rho (E+\omega )}.
\label{nor}
\end{equation}

\section{Results}
\label{R}

We have commented above on the form of the average level density given
in Eq.~(\ref{dos}); these comments will not be repeated here. Rather,
we focus attention on the normalized level--level correlation function
defined in Eq.~(\ref{nor}). With
all energies scaled in units of the GUE mean level spacing $d = \pi
\lambda / N$ at the center of the unit circle, we present in Figure 1
the function $C_{nor}$ versus $\omega / d$ for parameter values as 
given in the
caption. For comparison, we also show the standard GUE level--level
correlation function $\sin^2(2 \pi \omega / d_{eff}) / (2 \pi \omega /
d_{eff})^2$. We note that in the figure, we have chosen $\Gamma = 0.3
d$ so that differences between the two functions should be visible in
the wings. As anticipated above, we find that the two functions
nearly coincide within the peak. While the GUE correlation function is
positive semidefinite everywhere, the function $C_{nor}$ oscillates in
the wings around zero.

In Figure 2, we show $C_{nor}$ for the case $\Gamma = 3 d$. Here, there
is no discernible difference between this function and the standard
GUE autocorrelation function $\sin^2(2 \pi \omega / d_{eff}) / (2 \pi 
\omega / d_{eff})^2$ {\it provided that the latter is scaled with}
$d_{eff}$. The second function in Figure 2 with the wide peak shows
the GUE correlation function evaluated without rescaling, i.e. the
function $\sin^2(2 \pi \omega / d / (2 \pi \omega / d)^2$. The
difference between both functions is substantial.

In summary, we have presented first analytical results for a 
random--matrix model describing wave propagation in a random medium 
with resonant scattering. We are
confident that this work will produce further interesting results.

\acknowledgements{V. K. gratefully acknowledges the support of 
a MINERVA Fellowship. 
B. E. wishes to thank Prof. A. Nourreddine for valuable discussions.}

\begin{figure}
\setlength{\unitlength}{0.240900pt}
\ifx\plotpoint\undefined\newsavebox{\plotpoint}\fi
\sbox{\plotpoint}{\rule[-0.200pt]{0.400pt}{0.400pt}}%

\label{fig1}
\end{figure}
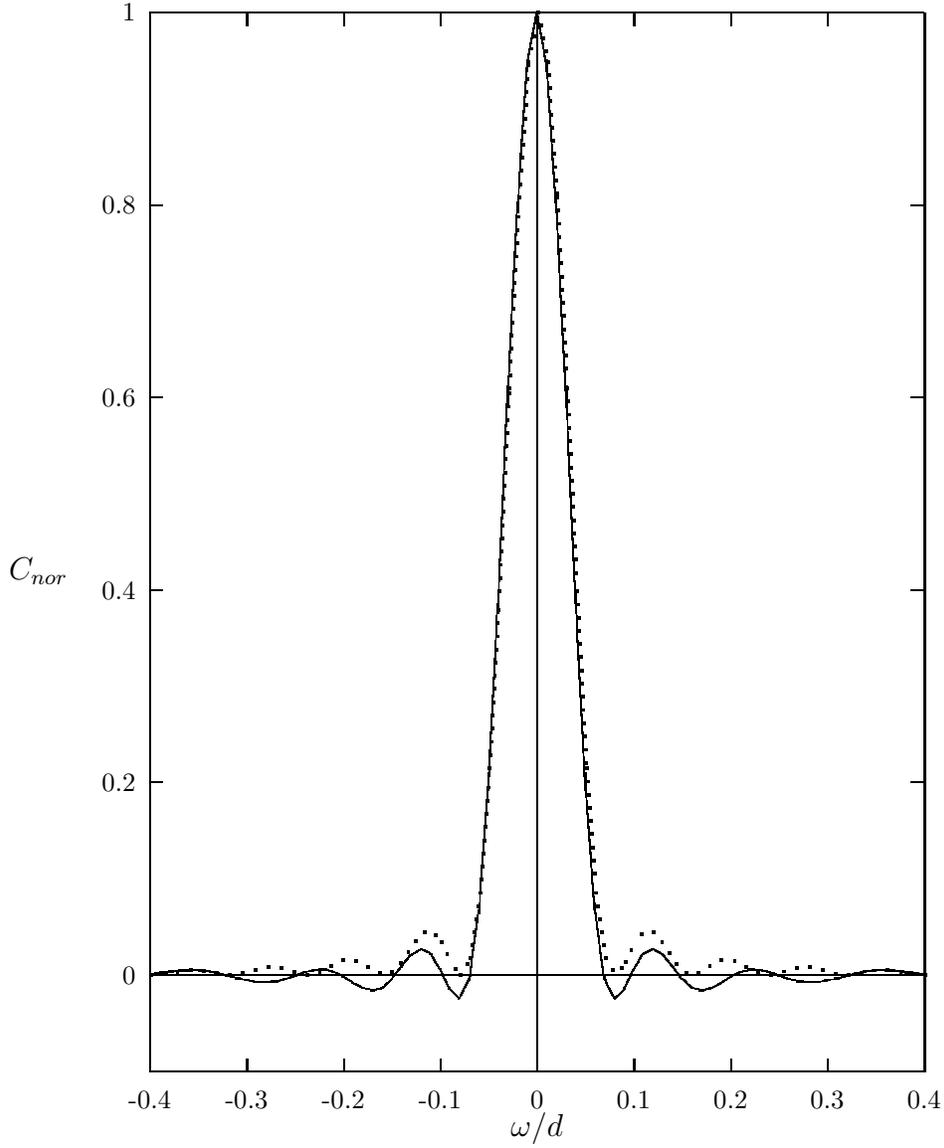
Fig.1. The normalized level--level correlation function of Eq.~(8) 
for $m=5$, $\Gamma =0.3d$, $E=E_1=0$ (full line) and
$\sin^2(2\pi\omega /d_{eff})/(2\pi\omega /d_{eff})^2$
 with 
$d_{eff}=d/(1+m/(\pi\Gamma ))$ (dotted line).

\begin{figure}
\setlength{\unitlength}{0.240900pt}
\ifx\plotpoint\undefined\newsavebox{\plotpoint}\fi
\sbox{\plotpoint}{\rule[-0.200pt]{0.400pt}{0.400pt}}%
%
\label{fig2}
\end{figure}
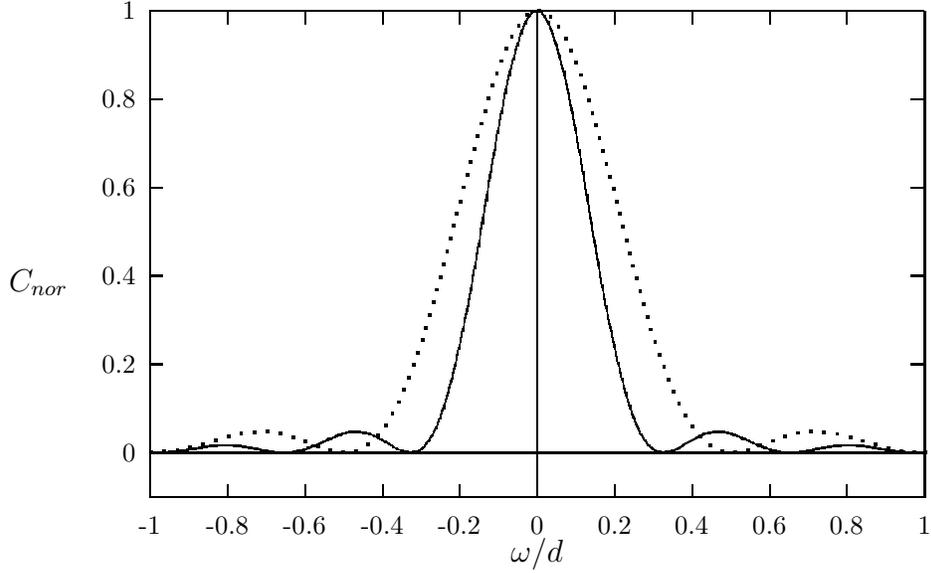
Fig.2. The normalized level--level correlation function of Eq.~(8) 
for $m=5$, $\Gamma=3d$, $E=E_1=0$
(which coincides almost exactly with 
$\sin^2(2\pi\omega /d_{eff})/(2\pi\omega /
d_{eff})^2$ with $d_{eff}=d/(1+m/(\pi\Gamma ))$) (full line) and 
the standard normalized GUE correlation function 
$\sin^2(2\pi\omega /d)/(2\pi\omega / d)^2$ (dotted line).

\end{document}